\documentclass[10pt]{iopart}

\newcommand{\id}{\mathbbm{1}}
\newcommand{\W}{{\mathcal W}}
\newcommand{\beq}{\begin{eqnarray}}
\newcommand{\eeq}{\end{eqnarray}}

\usepackage{bbm}
\usepackage{graphicx}

\begin{document}

\title{Mutually Unbiased Bases and Bound Entanglement}

\author{Beatrix C. Hiesmayr}

\address{University of Vienna, Faculty of Physics, Boltzmanngasse 5, 1090 Vienna, Austria.}
\ead{Beatrix.Hiesmayr@univie.ac.at}

\author{Wolfgang L\"offler}

\address{Leiden University, Quantum Optics \& Quantum Information, PO Box
9500, 2300 RA Leiden, Netherlands}

\begin{abstract}
In this contribution we relate two different key concepts: mutually unbiased bases (MUBs) and entanglement; in particular we focus on bound entanglement, i.e. highly mixed states which cannot be distilled by local operations and classical communications. For a certain class of states --for which the state-space forms a ``\textit{magic}'' simplex-- we analyze the set of bound entangled states detected by the MUB criterion for different dimensions $d$ and number of particles $n$. We find that the geometry is similar for different $d$ and $n$, consequently, the MUB criterion opens possibilities to investigate the typicality of PPT-bound and multipartite bound entanglement deeper and provides a simple experimentally feasible tool to detect bound entanglement.
\end{abstract}

\maketitle

\section{Introduction}

Mutually unbiased bases (MUBs) are a key concept in quantum science since they are intimately related to the nature of quantum information. Measurements made in one basis of a set of MUBs provide maximal uncertainty about the state if prepared in another basis from the same set. In quantum mechanics, the amount of information that can be extracted from a physical system is fundamentally limited. In this context, MUBs acquire a fundamental relevance since they serve as a powerful toolbox to explore generally Bohr's complementarity. Also in mathematical science MUBs are investigated since the problem of finding a complete set of MUBs for a given dimension is only solved for prime and power prime dimensions: the so called maximal set problem.

Key evidence of entanglement are correlations that are stronger than those possible in classical systems. In 1998 the Horodecki family predicted a new type of entanglement: bound entanglement~\cite{Horodecki}. These heavily mixed entangled states, from which no pure entanglement can be distilled applying local operations and allowing for classical communication, are puzzling since its role in Nature is not understood. Moreover, it is computationally hard to detect bound entanglement.

In this contribution we show a relation between these two basic concepts of quantum theory, in particular we show that bound entanglement can be detected in an experimentally feasible way via correlations obtained from a complete set of MUBs in prime and prime power dimensions. Last but not least we report on an experiment that via MUBs witnessed for the first time the generation of bound entangled bipartite photons~\cite{MUB1}.

\section{Definitions}

\textbf{Definition ``MUB'':}
Two orthonormal bases $\mathcal{B}_1=\{|i_1\rangle\}$ and $\mathcal{B}_2=\{|i_2\rangle\}$ in a $d$ dimensional Hilbert space $\mathcal{C}^d$ are called mutually unbiased if and only if
\begin{eqnarray}
|\langle i_1|j_2\rangle|^2&=&\frac{1}{d}\qquad\textrm{for all}\quad i,j=1,\dots,d\;.
\end{eqnarray}

In prime and prime power dimensions complete sets of $(d+1)$ MUBs exist. In composite dimensions such as $d=6,10,12,\dots$ the question of whether complete sets of $(d+1)$ MUBs exist remains essentially open.

\noindent\textbf{Definition ``Partial Separability'':}
A pure $n$-partite quantum state $|\Psi_{(k-sep)}\rangle$ is defined to be $k$-separable if and only if it
can be written as a product of $k$ states $\psi_i$ ($i=1,\dots k$) each one living on a different and non-overlapped Hilbert subspace
\begin{equation}
 \label{k-sep1}
 |\Psi_{(k-sep)}\rangle= |\psi_1\rangle \otimes |\psi_2\rangle \otimes \ldots |\psi_k\rangle\quad\textrm{with}\quad k\leq n \;.
\end{equation}
This concept can be straightforwardly extended to mixed states. A mixed $n$-partite quantum state is defined to be $k$-separable if and only if it has a decomposition into $k$-separable pure states:
\begin{eqnarray}
 \label{k-sep2}
 \rho_{(k-sep)}&=&\sum_i p_i |\Psi_{(k-sep)}^i\rangle\langle \Psi_{(k-sep)}^i|\;\textrm{with}\; p_i\geq 0\;\textrm{and}\; \sum_i p_i=1\;.
\end{eqnarray}

The $k$-separability is in literature sometimes also referred to as $k$-producibility or depth of entanglement. From the above definition we
immediately obtain that if a state is $k$-separable, it is automatically also $k'$-separable for all $k'< k$. In particular, an $n$-partite state
is called fully separable, if it is $n$-separable, whereas it is called genuinely $n$-partite entangled, if and only if it is not
biseparable ($2$-separable). If neither of these is the case, the
state is called partially entangled or partially separable.

While the above intuitive definition has been shown to provide a proper characterization of multipartite systems, it is far from straightforward to find out for a given state whether it is $k$-separable or not. One immediately recognizes this since a $k$-separable mixed state may be $k$-separable under different partitions.
%E.g., a tripartite biseparable state can be given by
%\begin{eqnarray}
%&&\rho_{(biseparable)}=\nonumber\\
%&&\sum p_i \rho_{AB}^i\otimes\rho_C^i+\sum q_j \rho_{AC}^j\otimes\rho_B^j+\sum r_k \rho_{BC}^k\otimes\rho_A^k\nonumber\\
%&&\textrm{with}\quad p_i,q_j,r_k\geq 0\quad\textrm{and}\quad \sum_i p_i+q_j+r_k=1\;,
%\end{eqnarray}
%where $\rho_{xy}$ describe bipartite states. The set of biseparable states is the convex hull of the states
%that are separable for a fixed bipartition.
From a physical point of view the generation of fully separable or biseparable states does not
require interaction of all parties. Moreover, from the quantum information theoretic and operational perspective genuine multipartite entangled
states allow for new applications such as quantum secret sharing (i.e. by distributing a secret over many different parties the genuine
multipartite entanglement assures security against eavesdropping~\cite{SHH,HHB}).

\noindent\textbf{Definition ``A \textit{Magic} Simplex For Bipartite and Multipartite Qudits''}

Let us start with the case of bipartite qudits. The first thing to note is that starting from any Bell state, the remaining $d^2-1$ orthogonal Bell states can be constructed by acting with Weyl operators in one subsystem, i.e.
\begin{eqnarray}
|\Phi^+\rangle &:=&\sum_{i=0}^{d-1}|ii\rangle\;,\quad
P_{0,0}\;{:=}\;|\Phi^+\rangle\langle\Phi^+|\\
P_{k,l}&:=&\id_d\otimes {\mathrm W_{k,l}}\;P_{0,0}\;\id_d\otimes
{\mathrm W}^\dagger_{k,l}
\end{eqnarray}
where the $\mathrm W_{k,l}$ are the Weyl operators defined by
\begin{eqnarray}
{\mathrm W}_{k,l}|s\rangle\;=\;
w^{k(s-l)}\;|s-l\rangle\qquad\textrm{with}\quad w\;=\;e^{2\pi
i/d}\;.
\end{eqnarray}
The so obtained $d^2$ orthonormal  Bell states  $P_{i,j}$ can be considered as the vertices of a real $d^2-1$ dimensional ``\textit{magic}'' simplex, i.e. considering all states that are convex combinations of these Bell states (Refs.~\cite{BHN1,BHN2})
\begin{eqnarray}
\W:=\left\lbrace\sum_{k,l=0}^{d-1} c_{k,l}\; P_{k,l}\;|\;c_{k,l}\geq
0,\;\sum_{k,l=0}^{d-1} c_{k,l}=1\right\rbrace\;.
\end{eqnarray}
One main property of this class of states is that tracing out one particle results in a
maximally mixed state, i.e. all states are locally maximally mixed.

For $n$ pairs of qudit states we can define a similar simplex by
\begin{eqnarray}
\mathcal{W}^{\otimes n}&:=&\left\lbrace\sum_{k,l=0}^{d-1}c_{k,l}\;\tilde{P}_{k,l}|c_{k,l}\geq0,\sum_{k,l=0}^{d-1} c_{k,l}=1\right\rbrace\;.
\end{eqnarray}
Also in this case, the $d^{2}$ Bell-type vertex states $\tilde{P}_{k,l}$
are obtained by applying a Weyl operator in one subsystem to $\tilde{P}_{0,0}=\frac{1}{d^{2}}\sum P_{k,l}^{\otimes n}$. For $d=2$ the vertex states are the famous Smolin states~\cite{Smolin} (certain mixtures of GHZ states) which are shown to be multipartite bound entangled.

In Ref.~\cite{HuberHiesmayr} it was proven that for any $n$ the geometrical properties concerning separability and mixedness are the same. In particular the authors showed that any entanglement witness for $n\geq 2$ can be reduced to the one obtained for a single pair (i.e. in $\W$). Therefore, all geometrical properties concerning full separability or positivity under partial transposition (PPT) can be derived considering the much easier case of a single pair (bipartite qudits). The geometry of entanglement visualized in Fig.~\ref{figbound} refers to states embedded in $\W$ as well as $\mathcal{W}^{\otimes n}$ for any $n\geq 2$.

\section{The MUB Criterion Detecting Bipartite or Multipartite Entanglement}

Consider the following scenario of a source producing two-qudit states $\rho\in\mathcal{C}^{d\times d}$, namely quantum states with $d$ degrees of freedom per particle. Both experimenters, Alice and Bob, can choose among $m$ different observables. \textit{What is the best strategy for Alice and Bob to detect the inseparability?}
The most striking difference between entanglement and separability are revealed by correlations in different basis choices. In particular, if Alice and Bob choose among $m$ different mutually unbiased bases, then the correlation functions $\mathcal{C}$ are bounded for all separable states by~\cite{MUB2} \begin{eqnarray}\label{eq:mubbound1}
I_{m}:=\sum_{s=1}^{m}C_{A_{s},B_{s}}\stackrel{\forall\textrm{separable states}}{\leq} 1+(m-1)\frac{1}{d}\stackrel{m=d+1}{=}2\;,
\end{eqnarray}
where $s$ denotes a MUB and $A/B$ the corresponding observables of Alice/Bob and $C$ is the correlation function -- a mutual predictability, i.e. the sum over all joint probabilities when both obtain the same measurement result: $C_{A_{s},B_{s}}:=\sum_{i=0}^{d-1}P_{A_{s},B_{s}}(i_{s},i_{s})$. Note that the labeling of the $d$ outcomes is not physical and therefore, can be chosen such that $I_{m}$ becomes maximal. It is not only an experimenter-friendly and powerful expression to test for entanglement in bipartite and multipartite systems, however, surprisingly it detects also bound entanglement~\cite{MUB1}, which we will elaborate further in this contribution.
Let us also remark that $I_{d+1}$ implies that if more than $d+1$ MUBs exist then we would have a conflict with the separability~\cite{MUB2}.

\section{Bound Entanglement in the Magic Simplex Detected by MUBs}

\begin{figure*}
\begin{center}
(a)\includegraphics[width=2.7cm]{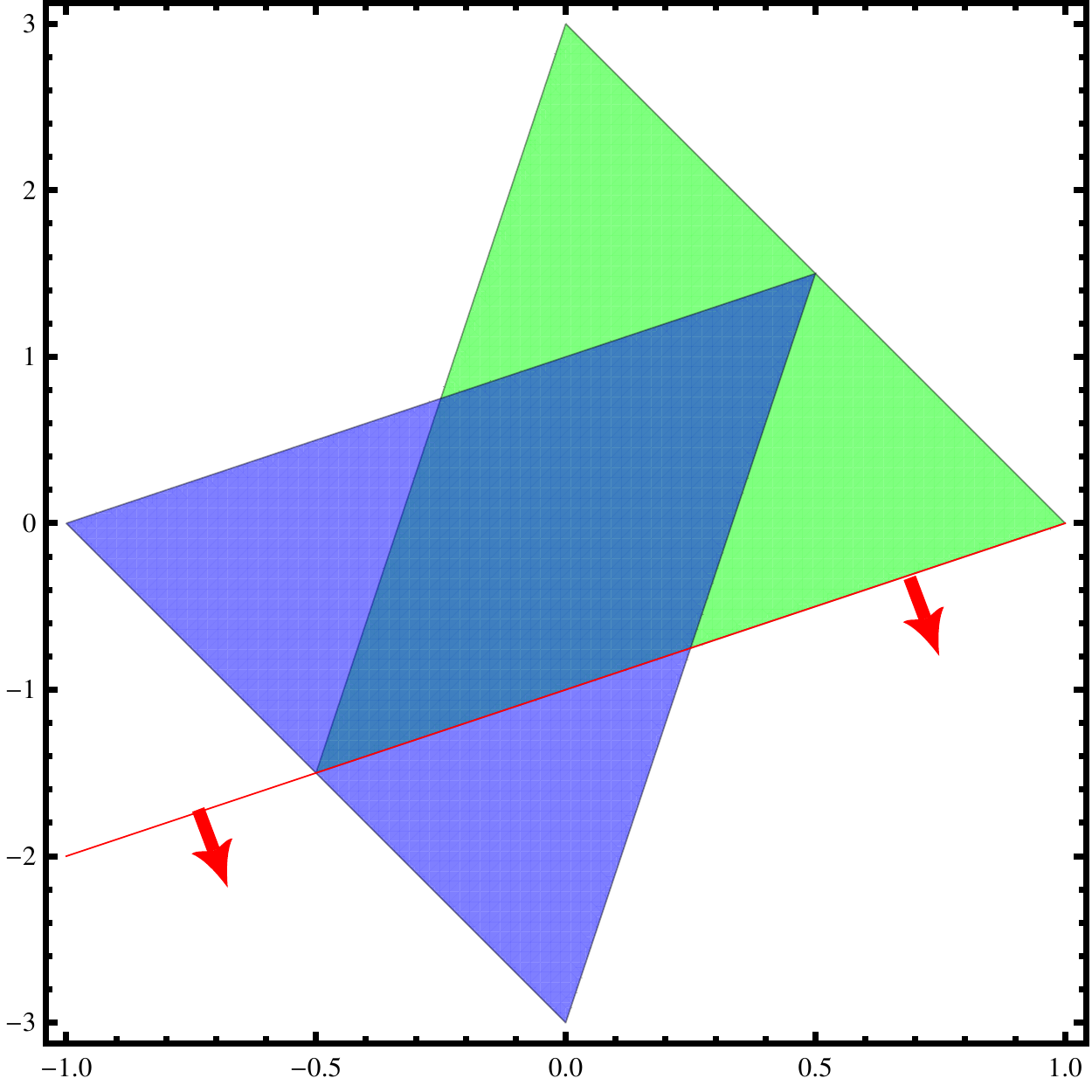}
(b)\includegraphics[width=2.7cm]{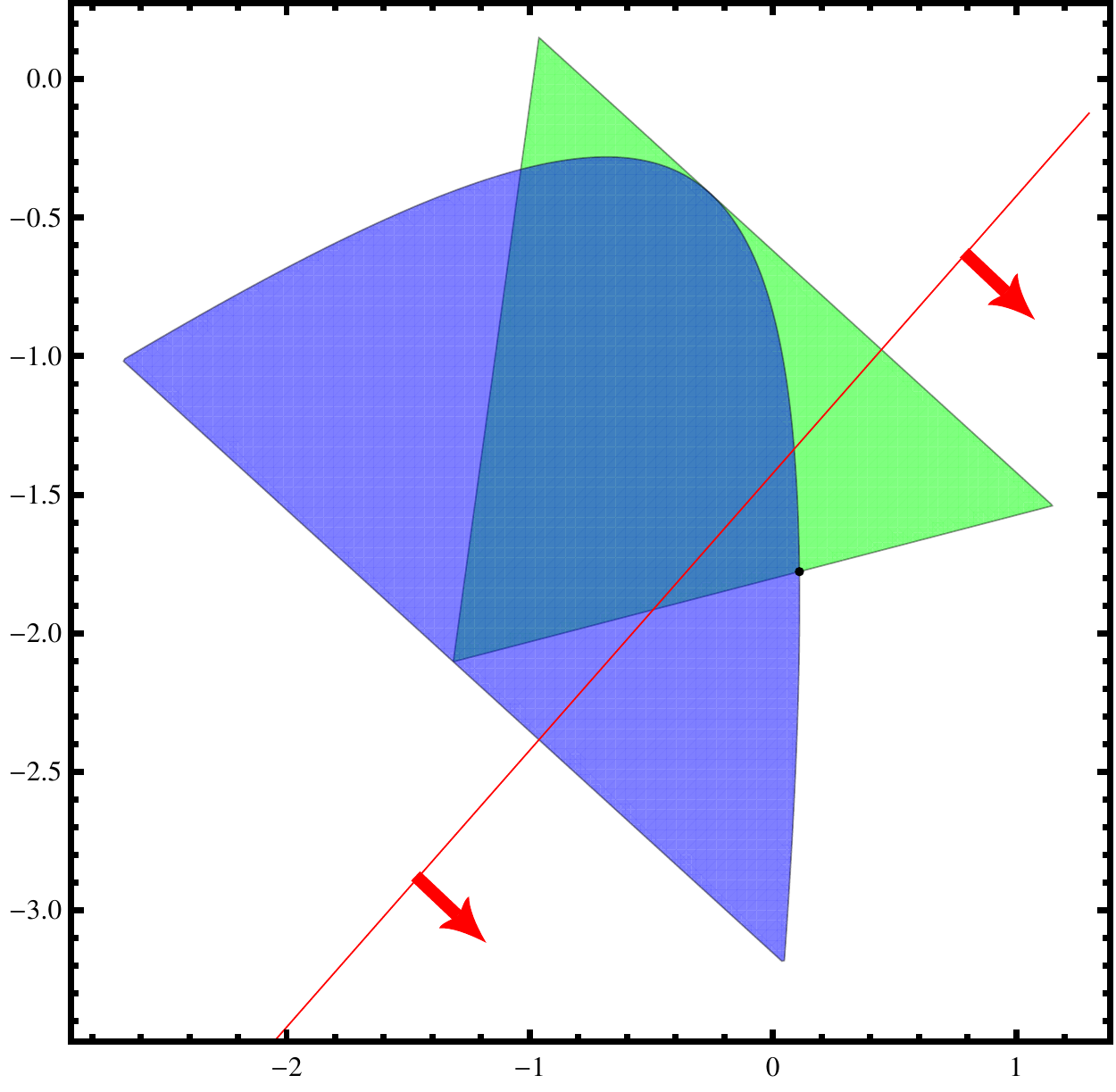}
(c)\includegraphics[width=2.7cm]{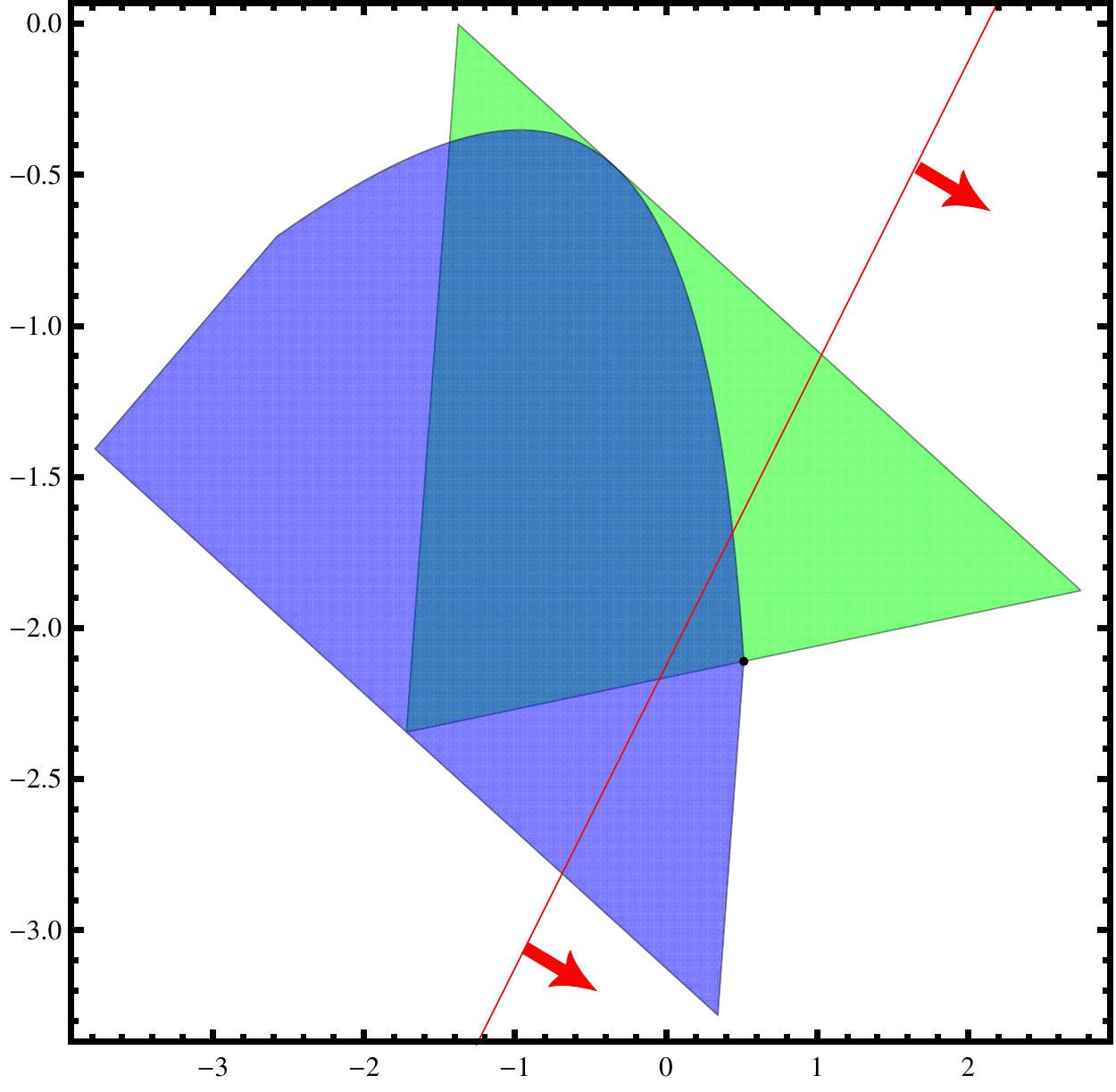}
(d)\includegraphics[width=2.7cm]{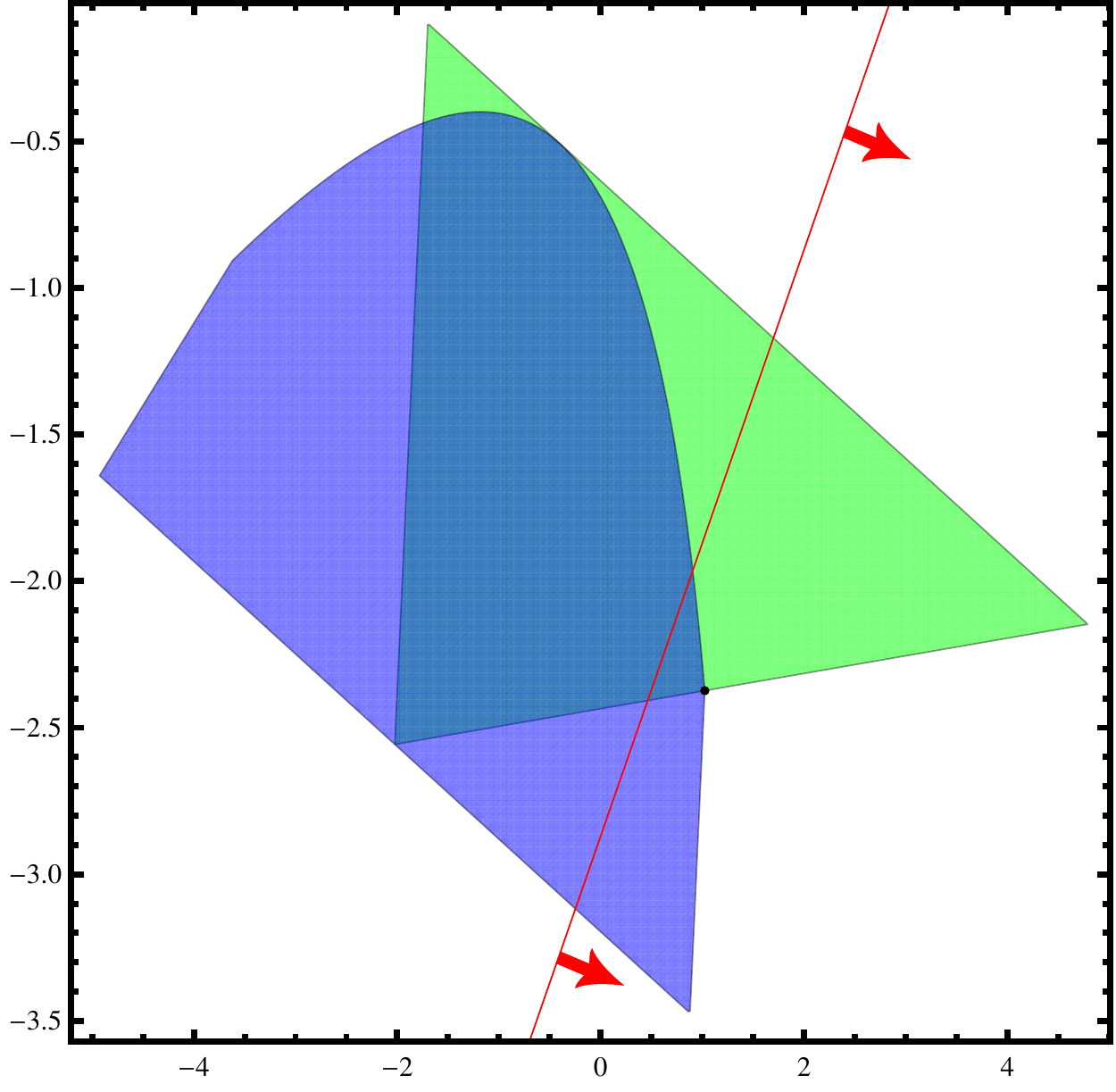}
(e)\includegraphics[width=2.7cm]{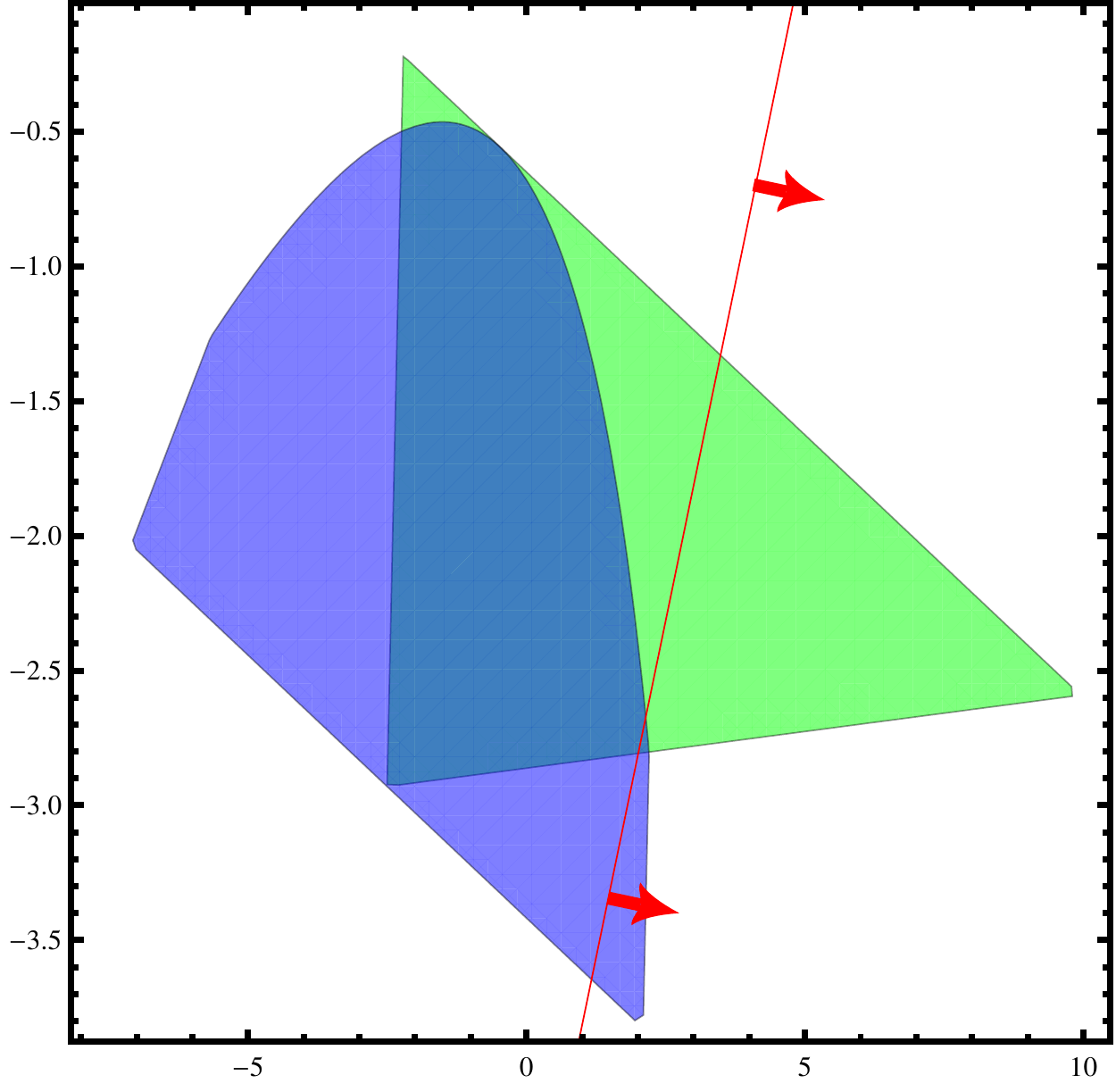}
(f)\includegraphics[width=2.7cm]{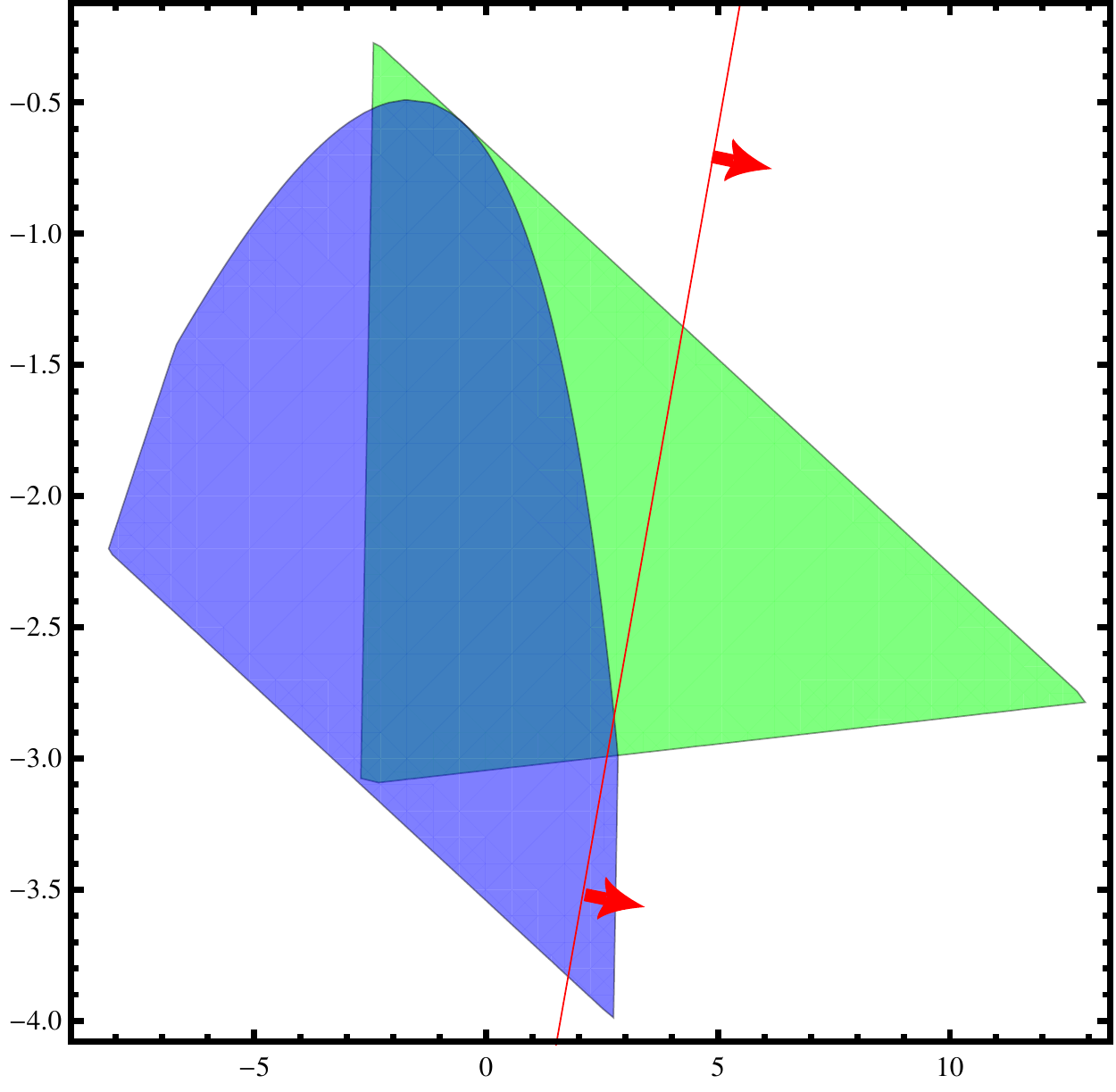}
(g)\includegraphics[width=2.7cm]{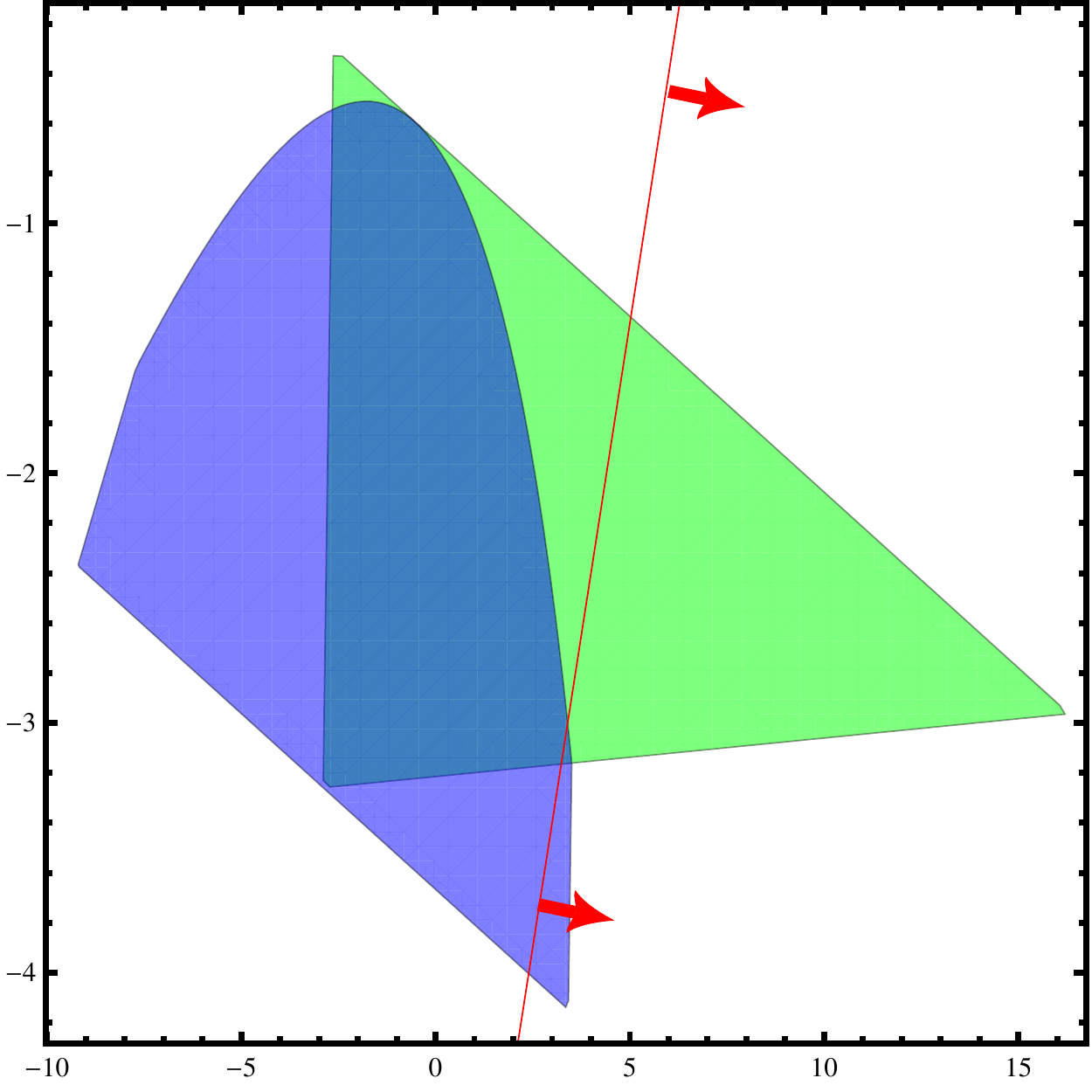}
\end{center}
\caption{Geometrical illustration of a slice through the magic simplex for bipartite and multipartite states: (a-g) present the state space of dimensions $d=2,3,4,5,7,8,9$, respectively. On the $x/y$-axis, the parameter $q_1/q_2$ of $\rho[d]$ or $\tilde{\rho}[d]$ are shown, respectively ($q_3,q$ fixed). The (green) triangle corresponds to the positivity condition whereas the curved (blue) region corresponds to states which are PPT. All states at the right hand side of the (red) line corresponds to states which are detected by the MUB criterion to be entangled $I_{d+1}>2$. Consequently, where the green and blue area overlap and $I_{d+1}>2$ the points correspond to bound entangled states. For $d=2$ no bound entanglement exist.}
\label{figbound}
\end{figure*}

Let us consider a class of states embedded in $\W$ (and by changing $P_{k,l}\longrightarrow \tilde{P}_{k,l}$ these states are embedded in  $\mathcal{W}^{\otimes n}$)
\begin{eqnarray}
\rho[d]&=&(1-\frac{q_1}{d^2-(d+1)}-\frac{q_2}{d+1}-q_3-(d-3)q)\frac{1}{{d^2}}\mathbbm{1}_{d^2}\nonumber\\
&&+ \frac{q_1}{d^2-(d+1)}P_{0,0}+\frac{q_2}{(d+1)(d-1)}\sum_{i=1}^{d-1} P_{i,0}\nonumber\\
&&+\frac{q_3}{d} \sum_{i=0}^{d-1} P_{i,1}+ \frac{q}{d}\sum_{z=2}^{d-2} \sum_{i=0}^{d-1} P_{i,z}\nonumber\\
\tilde{\rho}[d]&=&\quad P_{k,l}\longrightarrow \tilde{P}_{k,l}
\end{eqnarray}
The parameterisation is chosen such that the criterion $I_m$ obtains a simple form.

The first thing to note is that we found that PPT-bound entangled states of the above defined class could only be detected by $I_m$ if one considers a complete set of MUBs, i.e. $m=d+1$. We proved this for all prime and prime power dimensions up to $d=9$. In dimension $d=6$ only three MUBs are known, using them we do not detect any bound entanglement. This suggest that maximal complementarity, i.e. $m=d+1$, is needed for detecting bound entanglement.

In Fig.~\ref{figbound} we graphically summarize the geometry via slices through $\W$ or  $\mathcal{W}^{\otimes n}$, respectively. The parameters $q_3,q$ of states $\rho[d]$ (or $\tilde{\rho}[d]$) are chosen such that the MUB criterion $I_{d+1}$ is extreme, i.e. we find
\begin{eqnarray}
\min_{q_{i},\rho_{d}\geq0,\rho_{d}^{T_{A}}\geq0}2-I_{d+1}[\rho_{d}]=\left\lbrace \begin{array}{c}
-0.15\,(d=3)\\
-0.125\,(d=4)\\
-0.106\,(d=5)\\
-0.081\,(d=7)\\
-0.073\,(d=8)\\
-0.067\,(d=9)
\end{array}\right.
\end{eqnarray}
Then the graphics of Fig.~\ref{figbound} show the variation over $q_1$ ($x$-axis) and $q_2$ ($y$-axis) for different dimensions. The positivity condition leads to a (green) triangle while the Horodecki-Peres criterion (PPT) leads to a curved (blue) region. Hence, where both regions overlap each point corresponds to a state which is positive under partial transposition. All points at the right hand side of the red line are detected by the MUB criterion $I_{d+1}$ to be entangled, i.e. are greater than $2$. Consequently, where all three regions overlap the states are bound entangled. Note that the graphics could be also interpreted differently: Starting from states that are defined by the curved (blue) area the positivity criterion under partial transposition corresponds to states represented by the (green) triangle.

For $d=2$ no bound entanglement is detected in agreement with the fact that for bipartite systems bound entanglement only exists for $d\geq3$. For the other dimensions we observe that the geometry of the class of states that we consider is similar, and that a large region of states are detected to be bound entangled by the MUB criterion.

Let us now discuss the multipartite case for $d=2$. This simplex was first considered in Ref.~\cite{HHHKS}. It was found that all states are $2n$–-partite entangled, in particular they are not bipartite
entangled. Moreover, the vertex states, the Smolin states (certain mixtures of GHZ-states), cannot be distilled by local parties; they are bound entangled. Note that this multipartite bound entanglement is considerably different from bipartite bound entanglement, since -- if two local parties join -- pure maximal entanglement can be distilled by local operations and classical communications, i.e. the multipartite entanglement is unlockable. Interestingly, the entangled states inside the simplex can be distilled to the vertex states. In Ref.~\cite{HuberHiesmayr} the $2n$-partite states with $d=3$ degrees of freedom were considered and for a subset of states inside the simplex, a distillation protocol was found that transforms these states to the vertex states, which are unlockable bound entangled.

In summary, this means that while the basic geometric structure of separable and (PPT--bound) entangled states in the simplex remains unchanged with
$n$ (the pictures depict in Fig.~\ref{figbound} correspond to $\rho[d]$ and $\tilde{\rho}[d]$, respectively), the properties of the states change drastically. For $n\geq 2$ the states of the simplex are unlockable bound entangled since the (mixed) vertex states cannot be distilled, but if at least two local parties join the entanglement can be unlocked. In addition entangled states that are not $PPT$ inside the simplex may be distillable to any vertex state, whereas PPT--bound entangled states cannot be distilled at all.

In Ref.~\cite{MUB1} we have reported on an experiment with two photons, each one entangled in three different orbital angular momentum states. The correlations required for the MUB criterion $I_{d+1}$ were measured for $\rho[3]$ choosing proper parameters (Fig.\ref{figbound}(b)). A result above $2$ was recorded and thus proving the generation of entanglement. The positivity under partial transposition was proven via state tomography. Thus bipartite bound entanglement for qutrits was witnessed for the first time. Recent experiments in photonic multipartite systems~\cite{multibound1,multibound2} report the detection of a unlockable bound vertex states of four photons with $d=2$. In Ref.~\cite{QuesandaSanpera} it was reported that the Jaynes-Cummings model also exhibits bound entangled states.

Last but not least let us mention that not all bound entangled states in the magic simplex are detected by MUB criterion, e.g. those considered in Refs.\cite{BHN1,typicallybound}. This is also obvious since $I_{m}$ is a linear entanglement witness.

\section{Summary and Outlook}

We showed that the criterion based on mutually unbiased bases -- exploring Bohr's complementarity -- is a simple and experimentally feasible tool to detect PPT-bound entanglement in bipartite and multipartite states. In particular, the real dimensional ``magic'' simplex, a certain class of states, turned out to be helpful in comparing distinct dimensions and number of particles via the geometry. We find similar geometrical structures concerning the properties entanglement.

Since experimenters control more and more degrees of freedoms and number of particles our MUB criterion serves as a toolbox to reveal different entanglement features and may also contribute in solving the problem of the number of a complete set of MUBs.

\subsection{Acknowledgments}
BCH gratefully acknowledges the Austrian Research Fund FWF-P23627N16 and WL support by NWO.


\section*{References}
\begin{thebibliography}{10}

\bibitem{Horodecki}
M. Horodecki, P. Horodecki and R. Horodecki, % Mixed-State Entanglement and Distillation: Is there a “Bound”
%Entanglement in Nature?
Phys. Rev. Lett. 80, 5239 (1998).

\bibitem{MUB1}
B.C. Hiesmayr and W. L\"offler,
%Complementarity Reveals Bound Entanglement of Two Twisted Photons
New J. Phys. 15, 083036 (2013).

\bibitem{SHH}
St. Schauer, M. Huber, and B.C. Hiesmayr, Phys. Rev. A 82, 062311 (2010).

\bibitem{HHB}
M. Hillery, V. Bu\v{z}ek, and A. Berthiaume, Phys. Rev A  59, 1829 (1999).


\bibitem{BHN1}
B. Baumgartner, B.C. Hiesmayr and H. Narnhofer,
%The state space for two qutrits has a phase space structure in its core
Phys. Rev. A 74, 032327 (2006).

\bibitem{BHN2}
B. Baumgartner, B.C. Hiesmayr and H. Narnhofer,
%A special simplex in the state space for entangled qudits
J. Phys. A: Math. Theor. 40,  7919 (2007).




\bibitem{Smolin}
J.A. Smolin, Phys. Rev. A 63, 032112 (2001).

\bibitem{HuberHiesmayr}
B.C. Hiesmayr and M. Huber,
``\textit{Two distinct classes of bound entanglement: PPT-bound and `multi-particle'-bound}'', eprint: arXiv:0906.0238.

\bibitem{MUB2}
Ch. Spengler, M. Huber, St. Brierley, Th. Adaktylos, and B. C.
Hiesmayr, %Entanglement detection via mutually unbiased bases.
Phys. Rev. A \textbf{86}, 022311 (2012).


\bibitem{HHHKS}
B.C. Hiesmayr, F. Hipp, M. Huber, Ph. Krammer and Ch. Spengler,
%A simplex of bound entangled multipartite qubit states
Phys. Rev. A 78, 042327 (2008).

\bibitem{multibound1}
E. Amselem and M. Bourennane, %Experimental four-qubit bound entanglement.
Nature Phys. 5, 748 (2009).

\bibitem{multibound2}
J. Lavoie, R. Kaltenbaek, M. Piani and K.J. Resch,
%Experimental Bound Entanglement in a Four-Photon State.
Phys. Rev. Lett. 105, 130501 (2010).



\bibitem{QuesandaSanpera}
N. Quesada and A. Sanpera, ``\textit{Bound entanglement in the Jaynes Cummings model}'', preprint: arXiv:13052604.

\bibitem{typicallybound}
J. Bae, M. Tiersch, S. Sauer, F. de Melo, F. Mintert, B. C. Hiesmayr and A. Buchleitner,
%Detection and typicality of bound entangled states
Phys. Rev. A 80, 022317 (2009).

%\bibitem{book1} Goosens M, Rahtz S and Mittelbach F 1997 {\it The \LaTeX\ Graphics Companion\/}
%(Reading, MA: Addison-Wesley)
%\bibitem{eps} Reckdahl K 1997 {\it Using Imported Graphics in \LaTeX\ } (search CTAN for the file `epslatex.pdf')
\end{thebibliography}
\end{document}